\documentclass[twocolumn]{article}
\usepackage[a4paper, total={180mm, 243mm}]{geometry}
\usepackage[utf8]{inputenc}
\usepackage{amssymb}
\usepackage{bm}
\usepackage{soul}
\usepackage{lipsum}

\usepackage{graphicx}
\usepackage{hyperref}
\usepackage[
    backend=biber,
    style=nature,
    url=false,
    isbn=false,
    eprint=false,
    date=year,
    maxbibnames=99
]{biblatex}
\AtEveryBibitem{%
  \clearfield{note}%
}
\usepackage[T1]{fontenc}
\usepackage{siunitx}
\usepackage{upgreek}
\usepackage{titling}
\usepackage{circledsteps}   
\usepackage{subcaption}
\usepackage{mathtools}
\usepackage{bm}
\usepackage[inkscapeformat=pdf]{svg}

\usepackage[version=4]{mhchem}  
\usepackage[switch]{lineno} 
\modulolinenumbers[5]

\DeclareUnicodeCharacter{2009}{\,} 

\DeclareCaptionLabelSeparator{custom}{\textbar}
\DeclareCaptionFormat{custom}{\textsf{\textbf{#1 #2} \small #3}}
\newcommand\authormark[1]{\textsuperscript{#1}}

\graphicspath{{./figures}} 

\makeatletter
\newcommand{\showfontsize}{\f@size{} pt}
\makeatother

\title{$f$-2$f$ self-referencing for silicon nitride photonics \\ via a heterogeneously integrated lithium niobate layer}

\author{Weichen Fan\authormark{1}, Bastian Ruhnke\authormark{1}, Camiel Op de Beeck\authormark{2}, Giulio Tavani\authormark{2}, \\ Anton Vasiliev\authormark{2},  Franz Kärtner\authormark{1,3}, Tobias Herr\authormark{1,3,*}}

\date{%
    \small $^1$Deutsches Elektronen-Synchrotron DESY, Notkestr. 85, 22607 Hamburg, Germany \\
    \small $^2$LIGENTEC SA, EPFL Innovation Park, 1024 Ecublens, Switzerland\\
    \small $^3$Physics Department, Universität Hamburg UHH, Luruper Chaussee 149, 22607 Hamburg, Germany\\
    $^*$tobias.herr@desy.de \\
}

\begin{document}

\maketitle

\noindent \textbf{
Heterogeneous integration combines complementary material properties in a single photonic platform. Here, we demonstrate on-chip $f$-2$f$ self-referencing in a heterogeneously integrated \ce{Si3N4}/\ce{LiNbO3} platform, in which supercontinuum generation in a \ce{Si3N4} waveguide and second-harmonic generation in a \ce{LiNbO3} layer are spatially separated and linked by adiabatic escalator couplers. Pumped by a 1560 nm mode-locked laser with 75 pJ on-chip pulse energy, we detect the carrier-envelope offset frequency with 30 dB signal-to-noise ratio in 300 kHz resolution bandwidth. A comparative measurement confirms that the second harmonic originates in the \ce{LiNbO3} layer and indicates opportunity for further improvements through periodic poling. These results establish heterogeneous \ce{Si3N4}/\ce{LiNbO3} integration as a viable building block for self-referenced ultrafast pulse sources in low-loss \ce{Si3N4} photonics.}

\subsection*{Introduction}
Silicon nitride (\ce{Si3N4}) and lithium niobate (\ce{LiNbO3}) are versatile platforms for integrated nonlinear photonics. Both have a wide bandgap, a broad transparency window from visible to mid-infrared, and are increasingly available in foundry-based fabrication of integrated photonics. The \ce{Si3N4} platform stands out owing to ultralow propagation loss, enabling high-quality factor resonators, and efficient access to third-order nonlinear effects, including the generation of supercontinua \cite{brès2023SupercontinuumIntegratedPhotonicsa}, optical parametric amplification \cite{riemensberger2022PhotonicIntegratedContinuoustravellingwavea}, and frequency combs \cite{pasquazi2018MicrocombsNovelGeneration,herr2026FrequencyCombsCoherent} in high-Q microresonators. Low propagation losses also provide the basis for emerging chip-integrated mode-locked lasers and pulse amplifiers 
\cite{singh2020CWModelockedLasera, cuyvers2021LowNoiseHeterogeneous, singh2024SiliconPhotonicsbasedHighenergy, gaafar2024FemtosecondPulseAmplification, singh2025WattclassSiliconPhotonicsbased, qiu2026HighPulseEnergyIntegratedModeLocked, li2026FullyIntegratedDispersionmanaged, zeng2026IntegratedYtterbiumGain}. While propagation losses are typically higher in \ce{LiNbO3}, it offers (in contrast to \ce{Si3N4}) strong second-order nonlinearity enabling electro-optic modulation, and efficient frequency conversion from ultraviolet to mid-infrared \cite{wu2024VisibletoultravioletFrequencyCombb, ludwig2024UltravioletAstronomicalSpectrograph,  fan2025SpectralDynamicsBroadband, zhou2025QuadraticSupercontinuumGeneration, ludwig2026MidInfraredContinuaVia, li2026SegmentchirpedPeriodicallyPoleda}, which may be enhanced through quasi-phase matching via periodic poling \cite{zhu2021IntegratedPhotonicsThinfilma}. 
Hybrid approaches that combine \ce{Si3N4} and \ce{LiNbO3} offer a promising route to leverage the advantages of both materials in one platform. While cascading a \ce{LiNbO3} chip after a \ce{Si3N4} chip has demonstrated the potential \cite{kowligy2020MidinfraredFrequencyCombsa}, heterogeneous integration can further improve the compactness, robustness, and scalability. 

Recent studies have demonstrated heterogeneously integrated \ce{Si3N4}/\ce{LiNbO3} and similarly \ce{LiTaO3}/Si platforms at telecommunication wavelengths, enabling applications in optical ranging via tunable self-injection locking, high-speed data transmission, and electro-optic frequency comb generation \cite{chang2017HeterogeneousIntegrationLithiuma,churaev2023HeterogeneouslyIntegratedLithiuma,snigirev2023UltrafastTunableLasersa,niels2026HighspeedHeterogeneousLithiuma,cai2026HeterogeneouslyIntegratedLithiuma}. 
Extending heterogeneous integration beyond the telecommunication band to broadband nonlinear photonics is attractive; a prime application demanding such broadband operation is the detection of the carrier-envelope offset frequency ($f_\mathrm{ceo}$) of femtosecond lasers and frequency combs, which is essential to optical time and frequency metrology \cite{udem2002OpticalFrequencyMetrologya}. A commonly used method for the measurement of $f_\mathrm{ceo}$ (and self-referencing) is $f$-$2f$ beatnote detection, which involves interfering the spectrally broadened source spectrum (supercontinuum) with its spectrally overlapping second harmonic on a photo-detector.
Beyond traditional approaches involving second-harmonic generation in bulk crystals after supercontinuum generation in fibers, recent progress has realized on-chip self-referencing across multiple material platforms by simultaneous spectral broadening and harmonic generation \cite{carlson2017SelfreferencedFrequencyCombsa, hickstein2017UltrabroadbandSupercontinuumGenerationc, okawachi2018CarrierEnvelopeOffseta, yu2019CoherentTwooctavespanningSupercontinuum,okawachi2020ChipbasedSelfreferencingUsinga, 
obrzud2021StableCompactRFtooptical, ishizawa2022DirectF3fSelfreferencing, fan2024SupercontinuaIntegratedGalliumb, hamrouni2024PicojoulelevelSupercontinuumGenerationb, wu2024VisibletoultravioletFrequencyCombb, tang2025LithiumNiobateMicrowaveguidesa, volpini2026CladdingBasedGaN, chi2026ChipbasedF2fInterferometryb, tang2026OnChipAmplificationFreeFCEOb}, including in a uniform, heterogeneously integrated \ce{Si3N4}/\ce{LiNbO3} waveguide, in which supercontinuum and second harmonic were generated simultaneously~\cite{churaev2023HeterogeneouslyIntegratedLithiuma}.

In this work, we demonstrate on-chip $f$-$2f$ self-referencing in a heterogeneously integrated \ce{Si3N4}/\ce{LiNbO3} platform, where light transitions between waveguides ina thick, anomalous-dispersion \ce{Si3N4} layer for supercontinuum generation and waveguides in a thin \ce{Si3N4} layer that exhibit a strong mode overlap with a \ce{LiNbO3} layer for second harmonic generation.
Both layers are bridged by escalators with low insertion loss over a broad spectral range. This design enables independent waveguide structures for spectral broadening and second-harmonic generation and mitigates competition between $\chi^{(2)}$- and $\chi^{(3)}$-nonlinear processes. With an on-chip pump energy of 75~pJ we detect an $f_\mathrm{ceo}$ beatnote with 30~dB signal-to-noise ratio (SNR) at 300~kHz resolution bandwidth (RBW), sufficient for $f_\mathrm{ceo}$ stabilization. This result opens new avenues for fully integrated, self-referenced frequency combs with applications in precision metrology and spectroscopy.

\begin{figure}[ht!]
    \centering
    \includegraphics[width=0.9\linewidth]{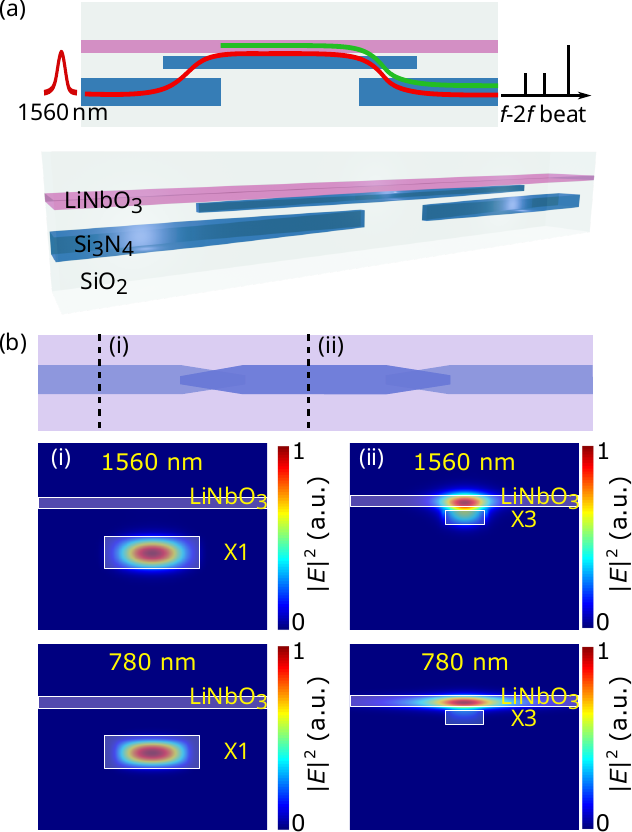}
    \caption{(a) Conceptual waveguide structure (not to scale) for $f$-$2f$ self-referencing on the heterogeneously integrated \ce{Si3N4}/\ce{LiNbO3} platform. Violet: \ce{LiNbO3}; Blue: \ce{Si3N4}; Faint cyan: \ce{SiO2}. The red line indicates the optical path of the pump light and the emerging supercontinuum, the green path indicates the second harmonic signal generated in the \ce{LiNbO3} layer; their interference and detection leads to the $f$-$2f$ beatnote signal.
    (b) Top-view of (a) and simulated profiles of the fundamental TE mode of 1560~nm and 780~nm wavelengths at the positions marked by (i) 3000~nm waveguide width in X1 layer, and (ii) 1200~nm waveguide width in X3/\ce{LiNbO3} layer, respectively.}
    \label{fig_concept}
\end{figure}

\subsection*{Design}

\begin{figure}[ht!]
    \centering
    \includegraphics[width=0.9\linewidth]{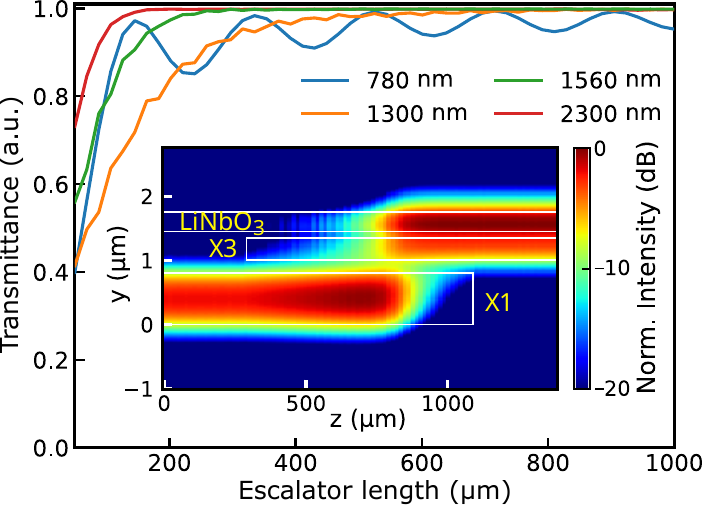}
    \caption{Simulated transmittance of 780~nm~(blue), 1300~nm~(orange), 1560~nm~(green), and 2300~nm~(red) wavelengths with varying escalator length. Inset shows the simulated propagation profile in the waveguide including \SI{800}{\micro m} long escalator with 1560~nm light source.}
    \label{fig_design}
\end{figure}

The heterogeneously integrated \ce{Si3N4}/\ce{LiNbO3} structure is fabricated by Ligentec in a wafer-level foundry process. It consists of two \ce{Si3N4} layers and one \ce{LiNbO3} layer, as illustrated in Fig.~\ref{fig_concept}a. 
The lower thick \ce{Si3N4} layer (X1 layer) allows anomalous dispersion for broad spectrum generation. The upper thin \ce{Si3N4} layer (X3 layer) confines the light together with the x-cut \ce{LiNbO3} slab that is design to cover the entire escalator and to generate second harmonic.
All three device layers are surrounded by \ce{SiO2} cladding. Figure~\ref{fig_concept}b shows the simulated mode profiles of the fundamental TE mode at 1560~nm and 780 nm wavelengths, where the waveguide width is 3000~nm in X1 layer and 1200~nm in X3 layer, respectively. The \ce{LiNbO3} slab has minor impact on the mode in the X1 waveguide core, however, with only an X3 waveguide core present, the optical mode shows large overlap with the \ce{LiNbO3} layer. Efficient coupling between X1 and X3 cores is achieved via escalators, where an adiabatic transition between both layers is achieved by tapers in opposite directions. The width in X1 layer is tapered down to 300~nm, while tapers in X3 layer increase from varying width to 1200~nm. The onset of the tapers is at the same longitudinal position as indicated in Figs.~\ref{fig_concept}a,b.
Figure~\ref{fig_design} shows the simulated transmittance of different wavelengths through one escalator as a function of the escalator length from \SI{50}{\micro m} to \SI{1000}{\micro m} in step of \SI{50}{\micro m} with 300~nm taper tip width in X3 layer (Eigen-Mode Expansion method, EME, Ansys Lumerical). The simulated transmittance indicates an adiabatic transition from the mode in X1 core to the mode in the X3 core across a spectral range from 1300~nm to 2300~nm with escalator length longer than \SI{700}{\micro m}. Importantly, the escalator also maintains a transmission at 780 nm wavelength above 90\% for a taper length of \SI{800}{\micro m} and longer; the latter is a crucial prerequisite for our $f$-$2f$ self-referencing scheme.

\subsection*{Experiment}
As a first measurement, we characterize the transmission performance of the escalator, with a broadband supercontinuum (SC) source spanning from visible to the mid-infrared range. After passing through a wire-grid polarizer and a notch-filter (for pump suppression), the SC source is coupled into the slow axis of a polarization maintaining single-mode (telecom wavelength) fiber via a reflective collimator. Polarization maintaining lensed fibers are used for both input and output coupling into the fundamental transverse electric (TE) mode of the X1 waveguides. 

\begin{figure}[ht!]
    \centering
    \includegraphics[width=0.9\linewidth]{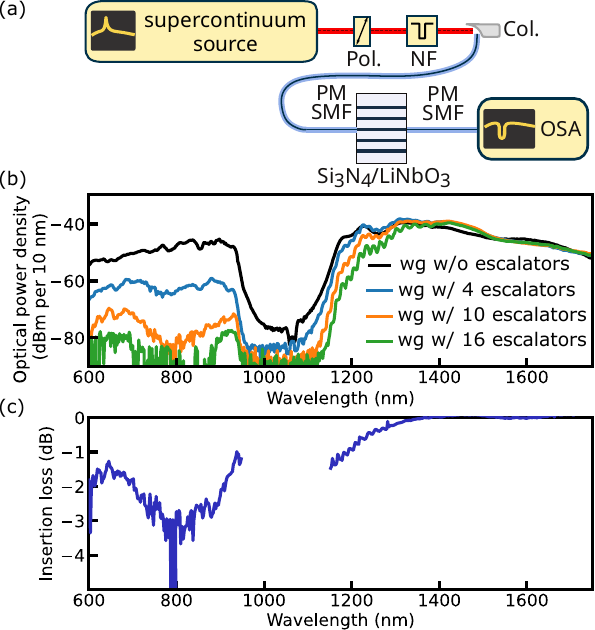}
    \caption{(a) Schematic setup for measuring the transmission of the escalator. Pol.: polarizer; NF: notch filter; Col.: collimator; PMSMF: polarization-maintaining single mode fiber; OSA: optical spectrum analyzer. (b) Measured transmission spectra of the reference waveguide without escalators (black), and the sample waveguides with 4 (blue), 10 (orange), and 16 (green) escalators, respectively. (c) Corresponding insertion loss of the escalator.
    }
    \label{fig_transmission}
\end{figure}

Waveguide transmission spectra are recorded for multiple waveguides containing 4, 10, or 16 escalators with alternating direction, where the even number ensures that the input and output are both in the X1 layer; a waveguide without escalators serves as a reference (Fig.~\ref{fig_transmission}b). Consistent with the simulation (Fig.~\ref{fig_design}), along one escalator the X1 core width changes from 3000~nm to 300~nm and the X3 core width increases from 300~nm to 1200~nm. The length of the escalators is \SI{800}{\micro m}. The reference waveguide maintains an X1 core width of 3000~nm. 
The loss per escalator is calculated from the transmission difference between the waveguides containing 4 and 10 escalators.
The results are in agreement with the simulation, indicating that the escalator has a broad transmission window. The loss per escalator is less than 1~dB for wavelengths exceeding 1200~nm, and below 4~dB at wavelengths below ca. 950~nm.

Next, we aim to detect the carrier-envelope offset frequency of an off-the-shelf 1560~nm mode-locked laser source with 100~MHz repetition rate and 100~fs pulse duration. For $f_\mathrm{ceo}$ signal generation we use a photonic chip, supporting a total straight propagation length of 15.83~mm. The initial 13.115~mm of propagation are in an X1 layer core of 2000~nm width, offering broadband weak anomalous dispersion around the pump as well as higher order dispersion for dispersive wave generation in the 700-800~nm wavelength range. This initial section is followed by an \SI{800}{\micro m}-long escalator to transition the light to the X3 core (with strong \ce{LiNbO3} mode overlap); here, the X3 taper starts with 900~nm width, which has negligible impact on insertion loss (extra 0.17~dB/<0.5~dB at 1560~nm/780~nm compared to an initial width of 300~nm), but permits a more compact escalator. After a propagation distance of \SI{300}{\micro m} for second harmonic generation, a reversed second escalator with the same structure is used to couple the light back to the X1 layer waveguide. Inverse tapers in the X1 layer are used for the coupling into and out from the chip. Input coupling of the mode-locked laser pulses occurs via a polarization maintaining lensed fiber, aligned to the fundamental TE mode of the waveguide (which is parallel to the optical axis of \ce{LiNbO3}). The coupling efficiency from the lensed fiber to the waveguide is estimated to be 25\% per facet, by measuring the throughput transmission of the corresponding reference waveguide (a 2000~nm wide waveguide in X1 layer)
with a continuous wave laser at 1560~nm. It indicates an on-chip energy of 75~pJ to drive the supercontinuum for a 300~pJ input energy before the chip.

\begin{figure}[ht!]
    \centering
    \includegraphics[width=\linewidth]{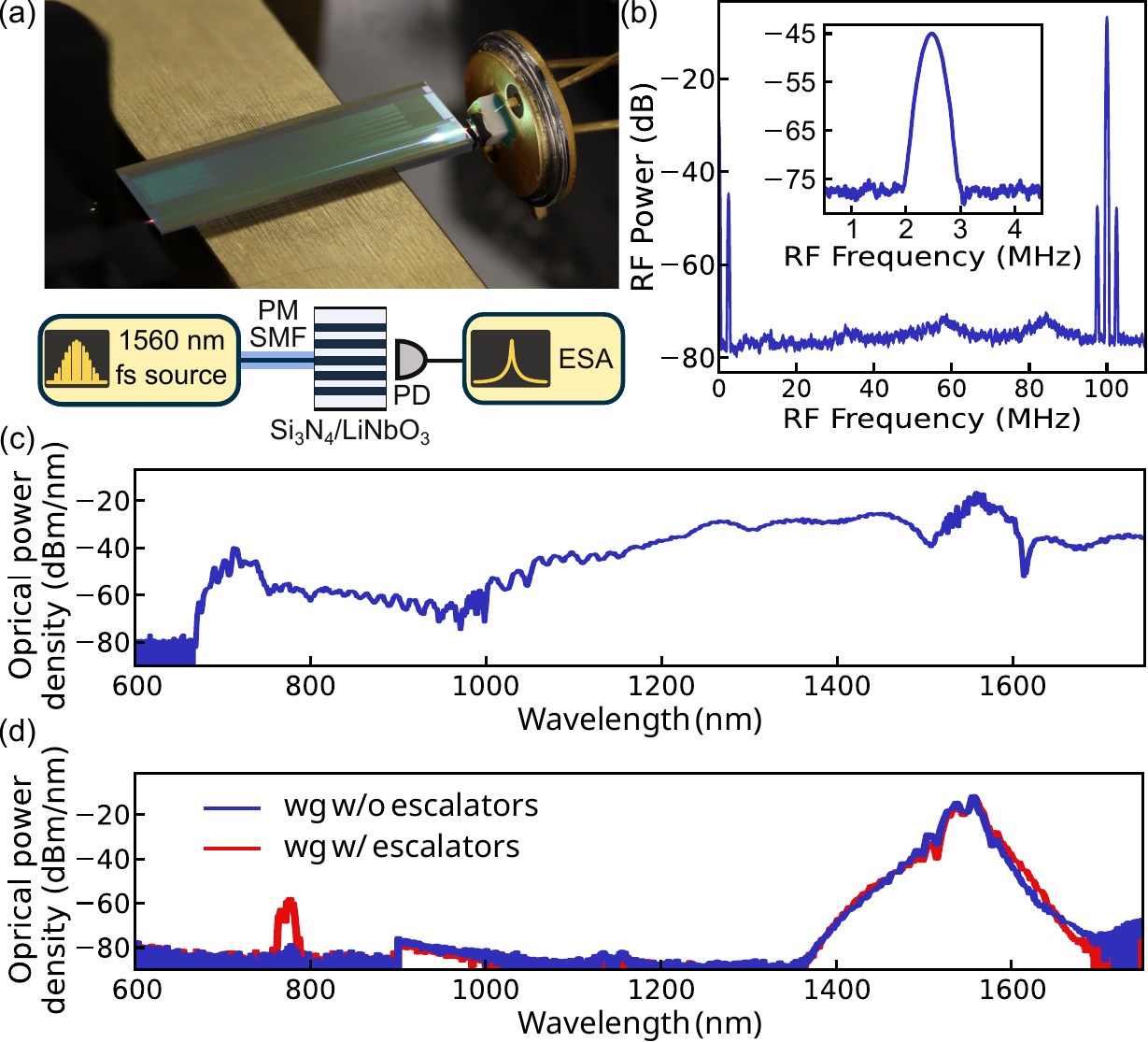}
    \caption{(a) A photo of the chip under test and the schematic setup for $f$-$2f$ beatnote measurement. PD: photodiode; ESA: electrical spectrum analyzer. (b) Measured RF beatnote with 300~kHz resolution bandwidth (RBW) and 3~kHz video bandwidth (VBW). Inset: magnified view of the $f_\mathrm{ceo}$ signal at 2.5~MHz. (c) Corresponding optical spectrum for the $f$-$2f$ beatnote measurement. (d) Output spectra from waveguides without the escalators (blue) and with escalators (red) pumped with temporally stretched pulses.}
    \label{fig_experiment}
\end{figure}

Figure~\ref{fig_experiment}c shows the broadband spectrum generated in the waveguides, which is collected by a multi-mode fluoride fiber and measured by an optical spectrum analyzer (OSA). The generated spectrum spans from 670~nm to beyond 1750~nm with a dispersive wave at 710~nm. 

For $f_\mathrm{ceo}$ detection a Si photodiode is directly placed at the output facet of the waveguide, as shown in Fig.~\ref{fig_experiment}a, and the RF signal is measured by an electrical spectrum analyzer (ESA). The corresponding RF beatnote is plotted in Fig.~\ref{fig_experiment}b, measured with 300~kHz RBW and 3~kHz video bandwidth (VBW). A pair of $f_\mathrm{ceo}$ signals can be seen in the RF spectrum, with 30~dB SNR, sufficient for stabilizing the $f_\mathrm{ceo}$ of the pump source.

To validate that the second harmonic giving rise to the $f_\mathrm{ceo}$ signal is generated by the \ce{LiNbO3} layer, we compare the output spectra of the waveguide with escalators and the reference waveguide without escalators under identical pumping conditions. To make the second harmonic spectrally distinguishable from the supercontinuum, the pump pulses are temporally stretched to ca. 700 fs (300 pJ pulse energy), suppressing supercontinuum formation while still permitting second-harmonic generation. The waveguide with escalators exhibits a second-harmonic signal 20~dB stronger than the reference (Fig.~\ref{fig_experiment}d), whose residual signal may originate from photogalvanic all-optical poling~\cite{nitiss2020BroadbandQuasiphasematchingDispersionengineered} of the \ce{Si3N4} or weak evanescent overlap with the \ce{LiNbO3} slab; the comparison demonstrates that the dominant contribution originates from the \SI{300}{\micro m}-long interaction with the \ce{LiNbO3} layer, despite the significant phase mismatch between pump and second harmonic (coherence length approximately \SI{2}{\micro m}). 

\subsection*{Discussion}
In summary, enabled by heterogeneous \ce{Si3N4}/\ce{LiNbO3} integration, we demonstrate a versatile integrated platform for self-referencing, directly compatible with low-loss \ce{Si3N4} photonics. With an on-chip pulse energy of 75~pJ, we observe an $f_\mathrm{ceo}$ signal with an SNR sufficient for phase stabilization (30~dB; RBW 300~kHz).
Considerable potential to lower the required pulse energy for $f$-$2f$ detection exists: First, a quasi-phase-matched \ce{LiNbO3} structure (e.g. via periodic poling) can drastically increase the second-harmonic conversion efficiency and hence the SNR of the $f$-$2f$ signal, permitting operation at reduced pulse energy. Second, the escalator design can be further optimized to reduce the insertion loss in the visible wavelength range. Third, for off-chip sources, the facet couplers can be optimized. Overall, our results position heterogeneously integrated \ce{Si3N4}/\ce{LiNbO3} as a viable broadband photonic building block in the emerging ecosystem of integrated femtosecond lasers and amplifiers~\cite{singh2020CWModelockedLasera, cuyvers2021LowNoiseHeterogeneous, singh2024SiliconPhotonicsbasedHighenergy, gaafar2024FemtosecondPulseAmplification, singh2025WattclassSiliconPhotonicsbased, qiu2026HighPulseEnergyIntegratedModeLocked, li2026FullyIntegratedDispersionmanaged, zeng2026IntegratedYtterbiumGain}, and contribute towards fully integrated self-referenced sources.

\subsection*{Acknowledgements}
This project has received funding from the European Innovation Council (EIC, grant agreements No 101046920 and No 101159229) and through the Helmholtz Young Investigators Group VH-NG-1404; the work was supported through the Maxwell computational resources operated at DESY.

\printbibliography

\end{document}